\documentclass[letterpaper]{llncs}

\usepackage{makeidx}  
\usepackage[nolist]{acronym}
\usepackage{icomma}
\usepackage{color}
\usepackage{array}
\usepackage{amsmath}
\usepackage{layouts}
\usepackage{booktabs}
\usepackage{mathptmx}
\usepackage{graphicx}
\graphicspath{{./}}
\usepackage{hyperref}
\usepackage{breakurl}
\newcommand{\bq}{\begin{equation}}
\newcommand{\eq}{\end{equation}}

\newcommand{\flop}{\mbox{flop}}

\newcommand{\cycles}{\mbox{cy}}

\newcommand{\GHZ}{\mbox{GHz}}

\newcommand{\GB}{\mbox{GB}}

\newcommand{\eos}{\;.}
\newcommand{\cma}{\;,}
\newcommand{\rlm}{roof{}line model}
\newcommand{\rl}{roof{}line}

\begin{document}

\begin{acronym}[DVFS]
    \acro{AGU}{address generation unit}
    \acro{AVX}{advanced vector extensions}
    \acro{CA}{cache agent}
    \acro{CL}{cache line}
    \acro{CoD}{cluster-on-die}
    \acro{DCT}{dynamic concurrency throttling}
    \acro{DIR}{directory}
    \acro{DP}{double precision}
    \acro{DP}{double precision}
    \acro{ECM}{execution-cache-memory}
    \acro{ES}{early snoop}
    \acro{FMA}{fused multiply-add}
    \acro{FP}{floating-point}
    \acro{HA}{home agent}
    \acro{HS}{home snoop}
    \acro{LFB}{line fill buffer}
    \acro{LLC}{last-level cache}
    \acro{MC}{memory controller}
    \acro{MSR}{model specific register}
    \acro{NT}{non-temporal}
    \acro{NUMA}{non-uniform memory access}
    \acro{OSB}{opportunistic snoop broadcast}
    \acro{RAPL}{running average power limit}
    \acro{SIMD}{single instruction, multiple data}
    \acro{SKU}{stock keeping unit}
    \acro{SP}{single precision}
    \acro{SP}{single precision}
    \acro{SSE}{streaming SIMD extensions}
    \acro{TDP}{thermal design power}
    \acro{UFS}{Uncore frequency scaling}
\end{acronym}

\frontmatter          
\pagestyle{headings}  
\addtocmark{Power Model} 

\mainmatter              
\title{On the accuracy and usefulness of analytic energy models for contemporary multicore processors}
\titlerunning{Accuracy and usefulness of analytic energy models}  
\author{Johannes Hofmann\inst{1} \and Georg Hager\inst{2} \and Dietmar Fey\inst{1}}
\authorrunning{Johannes Hofmann et al.} 
\tocauthor{Johannes Hofmann, Georg Hager}
\institute{Computer Architecture, University of Erlangen-Nuremberg, 91058 Erlangen, Germany,\\
\email{johannes.hofmann@fau.de, dietmar.fey@fau.de}
\and
Erlangen Regional Computing Center (RRZE), 91058 Erlangen, Germany,\\
\email{georg.hager@fau.de}}
\maketitle              

\begin{abstract}
This paper presents refinements to the execution-cache-memory
performance model and a previously published power model for multicore
processors. The combination of both enables a very accurate prediction
of performance and energy consumption of contemporary multicore processors
as a function of relevant parameters such as number of active cores as well as core
and Uncore frequencies. Model validation is performed
on the Sandy Bridge-EP and Broadwell-EP microarchitectures.
Production-related variations in chip quality are
demonstrated through a statistical analysis of the fit parameters
obtained on one hundred Broadwell-EP CPUs of the same model.  Insights
from the models are used to explain the performance- and
energy-related behavior of the processors for scalable as well as
saturating (i.e., memory-bound) codes. In the process we demonstrate the models' capability to
identify optimal operating points with respect to highest performance, lowest
energy-to-solution, and lowest energy-delay product and identify a set of
best practices for energy-efficient execution.
\keywords{Performance modeling, power modeling, energy modeling}
\end{abstract}

\section{Introduction}


The usefulness of analytic models for the performance and power
consumption of code running on modern processors is
undisputed. Here, ``analytic'' means a simplified description of
the interactions between software and hardware, simple enough
to identify relevant performance and energy issues but also elaborate
enough to be realistic at least in some important scenarios. There is
a large gray area between the extremes of modeling procedures: Purely
analytic, also called \emph{first-principles} or \emph{white-box}
models, try to start from known technical details of the hardware and
how the software executes, without additional phenomenological input
such as measured quantities or parameterized fit functions.  The other
end of the spectrum is set by \emph{black-box} models that can be
constructed from almost zero knowledge; 
measured runtime, hardware performance
metrics, power dissipation, etc., are used to identify crucial
influence factors for the metrics to be modeled.  One can
then use the ``trained'' system to predict properties of arbitrary
code, or play with parameters to explore design spaces. In either
case, the predictive power of the model enables insight beyond
what we would get by just running the code on the hardware at hand.

The power dissipation and energy consumption of HPC systems has become
a major concern. Developing a good understanding of
the mechanisms behind it and how code can be executed in the most
energy-efficient way is thus of great interest to
the community. It is
certainly out of the question that navigating the parameter space of
core count, clock frequencies, and (possibly) supply voltage will be
sufficient to meet the challenges of future top-tier parallel
computers is terms of power, but energy is still a major part
of the operating costs of HPC clusters. Moreover, there is a trend
to employ power capping in order to enable a more accurate tailoring
of the power supply to the needs of the machine, thereby saving
a lot of expenses in the infrastructure. Under such conditions,
letting code run ``cooler'' and knowing the energy vs.\ performance
tradeoffs will directly yield more science
(i.e., useful core hours) per dollar. 

This paper is concerned with core- and chip-level performance and
power models for Intel server CPUs. These models are precise enough to
yield quantitative predictions of energy consumption. In terms of
performance we rely on the execution-cache-memory (ECM) performance
model \cite{Hager:2012,sthw15} (of which the well-known \rlm\ is a
special case), which can deliver single-core and
chip-level runtime estimates for
loop-based code on multicore CPUs. A simple multicore
power model \cite{Hager:2012} serves as a starting point for energy
modeling. Both models are rather qualitative in nature; although
the ECM model is precise on the single core, it is over-optimistic
once the memory bandwidth starts saturating. The original power model
is very approximate and can only track the rough
energy consumption behavior of the processor. In this work we refine
both models to a point where the prediction accuracy for
performance and power dissipation, and thus also for energy
consumption, becomes unprecedented.  This comes at the price of
making the models more ``gray-box''-like in the above terminology,
i.e., they need more phenomenological input and fit parameters.
However, the actual choice of functional dependencies is still
motivated by white-box thinking.

This paper is organized as follows: The remainder of this section
describes related work and lists our new contributions. In
Section~\ref{sec:testbed} we describe the hardware and software
setup and our measurement methodology. Section~\ref{sec:ecmref}
refines the ECM performance model to yield more accurate
predictions for code near the bandwidth saturation point.
In Section~\ref{sec:powerref} we extend the simple multicore
power model by refining it for better baseline power prediction
and adapt it to the new Intel processors with dual clock
frequency domains (core and Uncore). Section~\ref{sec:energyval}
links the two models to validate the predicted energy consumption.
Motivated by the results we give some guidelines
for energy optimization in Section~\ref{sec:consequences}
and conclude the paper with an outlook to future work in
Section~\ref{sec:summary}.


\subsection{Related work}

Energy and performance models on the chip level have received
intense interest in the past decade. The \rlm\ \cite{roofline:2009}
is still the starting point for most code analysis activities, but
it lacks accuracy and predictive power on the single core and
for saturation behavior. The ECM model~\cite{Hager:2012,sthw15}
requires less phenomenological input but encompasses more
details of the underlying architecture than \rl, yielding
better results on the single core. In contrast to the
original ECM model we allow for latency penalty contributions
that depend on the memory bus utilization, making the model
accurate across the whole scaling curve.

Energy-performance tradeoffs have been studied since the power
envelope of processors became a major concern, but were only treated
phenomenologically~\cite{Freeh:2007,Song:2013}.  Rauber et
al.~\cite{Rauber:2012} show using a simple heuristic model
that the typical energy minimum versus clock speed observed for
scalable code can be derived analytically.  However, they do not have
a useful performance model and do not take saturation patterns due to
memory bandwidth exhaustion into account. Khabi et
al.~\cite{Khabi:2013} study the energy consumption of simple, scalable
compute kernels using a similar underlying power model, but they also
lack a performance prediction.  The energy model introduced by Hager
et al.~\cite{Hager:2012} includes performance saturation but is only
qualitative and thus allows only rough estimates of energy
consumption, due to the combined shortcomings of the underlying ECM
and power models.

Manufacturing variations of processors and their consequences have
been studied by several authors \cite{Inadomi:2015,Wilde:2015}, and we
do not add to their wisdom here; our contribution in this area is to show
the relation between fitting parameters for a specific specimen
and the ``batch,'' yielding insight about the usefulness of
a particular set of parameters.

\subsection{Contribution}

This paper makes the following contributions: We refine the ECM performance model
to accurately describe the saturation behavior of memory-bound loops across cores.
A previously published multicore power model is extended to include dual
clock domains (core and Uncore) and frequency- and core-dependent baseline power. The
achieved accuracy in predicting runtime, power, and energy (using both models
combined) with respect to core frequency, Uncore frequency, and number of
active cores is unprecedented. This is demonstrated with Intel Xeon Sandy Bridge and
Broadwell CPUs. We also identify which of the power model
parameters depend on the code and which do not. A statistical analysis
of the variation of power parameters due to production spread is given
for a batch of Intel ``Broadwell'' 10-core CPUs, setting the limits for
the generality of the power fit parameters. Finally, based on the
energy modeling results, we use Z-plots\footnote{We coined the name to honor
  a colleague that came up with the idea, not knowing that it had been
  around for some time. Lacking any better name, we stick to it here.}
to identify best practices for energy-efficient,
best-performance, and lowest-EDP (energy-delay product) execution of
scalable (\textsc{dgemm}) and saturating (\textsc{stream}) code, with
special emphasis on the Uncore clock of the Broadwell CPU, which we
identify as a crucial parameter in energy-aware computing.

%


\section{Testbed and methodology}
\label{sec:testbed}

All measurements were performed on one socket of
standard two-socket Intel Xeon servers.  A
summary of key specifications of the testbed processors is shown in
Table~\ref{tab:testbed}.  The Sandy Bridge-EP (SNB) and Broadwell-EP (BDW)
chips were selected for their
relevance in scientific computing. Along with their ``relatives,'' the Ivy
Bridge-EP (IVB) and Haswell-EP (HSW) microarchitectures, they make up more than 85\% of
the systems in the latest \textsc{top500} list published in November 2017.
\begin{table*}[!tb]
\centering
    \caption{\label{tab:testbed}Key specification of test bed machines.}
\begin{tabular}{l @{\hskip 0.3in} c @{\hskip 0.1in} c}
\toprule
Microarchitecture (Shorthand)   & Sandy Bridge-EP (SNB)                     & Broadwell-EP (BDW)                          \\
\midrule
Chip Model                      & Xeon E5-2680                              & E5-2697 v4                            \\
Thermal Design Power (TDP)      & 130\,W                                    & 145\,W                                \\
Supported core frequency range  & 1.2--2.7\,GHz                             & 1.2--2.3\,GHz                         \\
Supported Uncore frequency range& 1.2--2.7\,GHz                             & 1.2--2.8\,GHz                         \\
Cores/Threads                   & 8/16                                      & 18/36                                 \\
Core-private L1/L2 cache capacities  & 8$\times$32\,kB / 8$\times$256\,kB          & 18$\times$32\,kB / 18$\times$256\,kB    \\
Shared L3 cache capacity        & 20\,MB (8$\times$2.5\,MB)                 & 45\,MB (18$\times$2.5\,MB)            \\
Memory Configuration            & 4 ch. DDR3-1600                           & 4 ch. DDR4-2400                       \\
Theoretical Memory Bandwidth      & 51.2\,GB/s                                & 76.8\,GB/s                            \\
\bottomrule
\end{tabular}
\end{table*}

Apart from obvious advances over processor generations such as the increased
core count or microarchitectural improvements concerning SIMD ISA extensions,
major frequency-related changes were made on the HSW/BDW microarchitectures.
On the older SNB/IVB  microarchitectures the chip's Uncore\footnote{On Intel
processors, the term Uncore refers to all parts of the chip that are not part
of the core design, such as, e.g., shared last-level cache, ring interconnect,
and memory controllers.} was clocked at the same frequency as CPU cores. On
Haswell/Broadwell chips, separate clock domains are provided for CPU cores and the Uncore.
As will be demonstrated, the capability to run cores and the
Uncore at different clock speeds proves to be a distinguishing feature of the
newer designs that has significant impact on energy-efficient operation.

Processors featuring the new Uncore clock domain provide a feature called
\emph{Uncore frequency scaling} that allows chips to dynamically set the Uncore
frequency based on the workload. When this feature is disabled the Uncore
frequency is fixed at the maximum supported frequency. Although not
officially documented, a means to manually set the Uncore frequency via
a model specific register is supported by all HSW, BDW, and Skylake
processors; starting with version 4.3.0, the \texttt{likwid-setFrequencies}
tool from the \textsc{likwid} tool suite\footnote{\url{http://tiny.cc/LIKWID}}
provides a comfortable way to manually set the Uncore frequency.

Since previous investigations of the \ac{RAPL} interface indicate that data
provided by this interface is of high quality
\cite{Hackenberg:2015}, all power-related empirical data was
collected via \ac{RAPL} using \texttt{likwid-perfctr} (also from the
\textsc{likwid} tool suite).  Representatives from the classes of scalable and
saturating applications for which performance and energy behavior was investigated were
\textsc{dgemm} (from Intel's \textsc{MKL}, version 16.0.1) and the
\textsc{stream} triad pattern (executed in \texttt{likwid-bench}, again from
the \textsc{likwid} tool suite), respectively.  Variance of empirical
performance and power data was addressed by taking each measurement ten times;
afterwards, the coefficient of variation\footnote{The coefficient of variation
is used to measure the \emph{relative} variance of a sample. It is defined as
the ratio of the standard deviation $\sigma$ to the mean $\mu$ of a sample.}
was used to asses variance---which in no case was higher than 2\%, indicating that
variance is not a problem.


\section{Refined ECM performance model}\label{sec:ecmref}

The ECM model takes into account predictions of single-threaded in-core execution time
and data transfers through the complete cache hierarchy. These predictions
can be put together in different ways, depending on the CPU architecture.
On all recent Intel server microarchitectures it turns out that
the model yields the best predictions if one assumes no (temporal) overlap of data transfers
through the cache hierarchy and between the L1 cache and registers,
while in-core execution (such as
arithmetic) shows full overlap. Scalability is assumed to be perfect
until a bandwidth bottleneck is hit. A full account of the ECM model
would exceed the scope of this paper, so we refer to~\cite{sthw15}
for a recent discussion.

One of the known shortcomings of the ECM model is that it
is rather optimistic near  the saturation
point~\cite{Hager:2012,sthw15}, i.e., it overestimates 
performance when the memory interface is nearly saturated. There are
several possible explanations for this effect. For example, it is
documented that Intel's hardware prefetching mechanism reduces the prefetch
distance when the memory bus is near saturation~\cite{ia32opt:2016},
which leads to larger latencies for individual accesses, causing
an additional latency contribution to the data access time
in the model. Thus the assumption that the scaling is linear with
unchanged data delay contributions across all cores until
the bandwidth bottleneck is exhausted cannot be upheld in this
simple form. Based on this insight we
make the following additional assumptions about performance saturation:
\begin{figure}[tb]
    \centering
    \hspace*{\fill}
    \includegraphics*[scale=0.8]{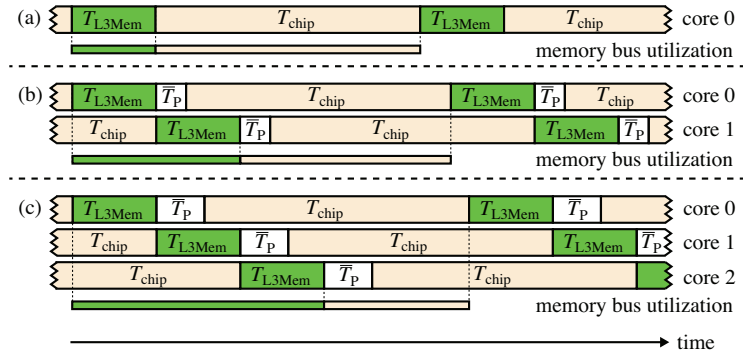}
    \hspace*{\fill}
    \caption{Visualization of memory bandwidth saturation under the refined
    ECM model. The white boxes show the average latency penalty $\bar
    T_\mathrm{P}$, which grows with with the utilization $u(n)$.}
    \label{ECM-model-refinement}
\end{figure}
\begin{figure}[tb]
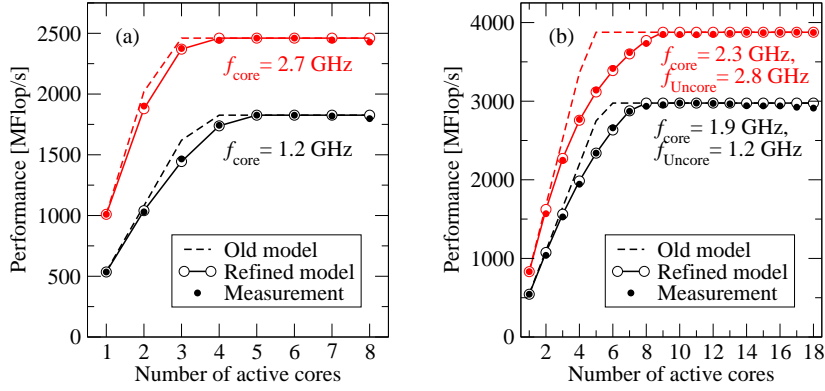

    \centering
    \hspace*{\fill}
    \includegraphics*[scale=0.70]{STREAM-multicore-SNB}
    \hfill
    \includegraphics*[scale=0.70]{STREAM-multicore-BDW}
    \hspace*{\fill}
    \caption{Comparison of original and refined ECM model multi-core estimates to
      empirical performance data on (a) SNB ($p_0=7.8\,\cycles$) (b) BDW
      ($p_0=5.2\,\cycles$) for the \textsc{stream}
    benchmark with a 16\,GB data set size.}
    \label{ECM-refinement}
\end{figure}
\begin{itemize}
\item Let $u(n)$ be the utilization of the memory interface with $n$
  active cores, i.e., the fraction of time in which the memory bus
  is actively transferring data. The plain ECM model predicts
  \bq
  u(1)=\frac{T_\mathrm{L3Mem}}{T_\mathrm{ECM}}=\frac{T_\mathrm{L3Mem}}{T_\mathrm{L3Mem}+T_\mathrm{chip}}
  \eq
  Here, $T_\mathrm{L3Mem}$ is the runtime contribution of L3-memory
  data transfers, and $T_\mathrm{chip}$ quantifies the data delay up
  to and including the L3 cache (see
  Figure~\ref{ECM-model-refinement}(a)).  No change to the model is
  necessary at this level.  
\item For $n>1$, the probability that a memory request initiated by a core
  hits a busy memory bus is proportional to the utilization of the bus
  caused by the $n-1$ remaining cores. If this happens, the core picks
  up an additional average latency penalty $\bar T_\mathrm{P}$
  (see Figures~\ref{ECM-model-refinement}(b) and (c))
  that is proportional to $(n-1)u(n-1)$:
  \bq
  \bar T_\mathrm{P}(n) = (n-1)u(n-1)p_0\eos
  \eq
  Here, $p_0$ is a free parameter that has to be fitted to the data.  
  Hence, we get a recursive formula for predicting the utilization:
  \bq\label{eq:memutil}
  u(n) = \min\left(1,\frac{nT_\mathrm{L3Mem}}{T_\mathrm{ECM}+\bar T_\mathrm{P}}\right) = \min\left(1,\frac{nT_\mathrm{L3Mem}}{T_\mathrm{ECM}+(n-1)u(n-1)p_0}\right)\eos
  \eq
  The penalty  increases
  with the number of cores and with the utilization, so it has the effect
  of delaying the bandwidth saturation.
\item The expected performance at $n$ cores is then
  $
  \pi(n) = u(n) \pi_\mathrm{BW}
  $,
  where $\pi_\mathrm{BW}$ is the bandwidth-bound performance limit as
  given, e.g., by the \rlm.
  If $u(n)<1$ for all $n\leq
  n_\mathrm{cores}$, where $n_\mathrm{cores}$ is the number of
  available cores connected to the memory interface (i.e.,
  in the ccNUMA domain), the code cannot saturate the memory bandwidth.
\end{itemize}
One phenomenological input for the ECM model is the saturated
memory bandwidth, which is typically determined by a streaming
benchmark. There is no straightforward way to derive the memory
bandwidth from core and Uncore clock frequencies, so we have to
measure its dependence on these parameters.
Figure~\ref{SNB-P_base-different-codes}(a) 
shows memory bandwidth versus (Un)core clock frequency on
BDW and SNB, respectively. The measured bandwidth at a particular
clock setting is then used together with the known memory traffic
to determine $T_\mathrm{L3Mem}$.

In Figure~\ref{ECM-refinement} we show comparisons between the original
scaling assumption (dashed lines) and our improved version (solid lines,
open circles) together with measured performance data (solid circles)
on the SNB and BDW chips. The agreement between the new model and the
measurement is striking. It is unclear and left to future work whether and
how $p_0$ depends on the code, e.g., the number of concurrent streams
or the potential cache reuse. Note that the single-core ECM model is
unchanged and does not require a latency correction.

Since there is no closed formula for the ECM-based runtime and
performance predictions due to the recursive nature of the utilization
ratio (\ref{eq:memutil}), setting up the model becomes more
complicated. We provide a python script for download that implements the
improved multi-core model (and also the power model
described in the following section) at \url{http://tiny.cc/hbpmpy}.
The single-core model can either be
constructed manually or via the open-source Kerncraft
tool~\cite{Hammer:2017}, which can automatically derive the ECM and
\rl\ models from C source and architectural information.

%
%



\section{Refined power model}\label{sec:powerref}

\begin{figure}[tbp]
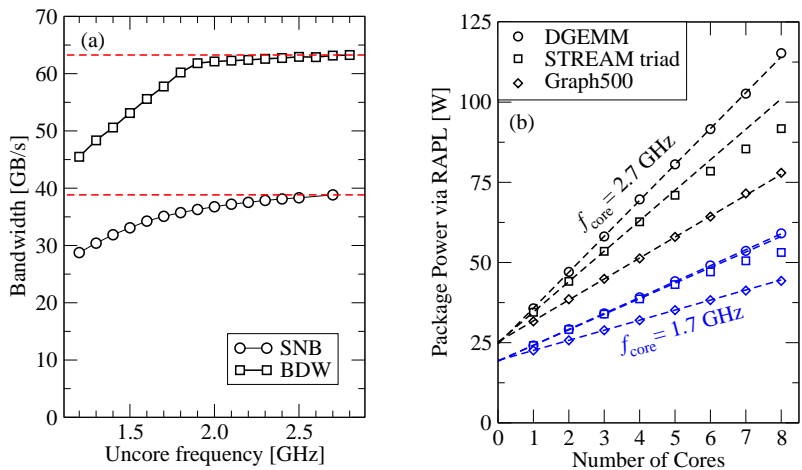

    \centering
    \hspace*{\fill}
    \includegraphics*[scale=0.70]{broadep2-uncore-bw}
    \hfill
    \includegraphics*[scale=0.70]{SNB-P_base-different-codes}
    \hspace*{\fill}
    \caption{(a) Maximum memory bandwidth (measured with \textsc{stream})
    versus (Un)core clock frequency on SNB and BDW.  (b) Package power
    consumption for \textsc{dgemm} ($N=20,000$), \textsc{stream}
    triad (16\,\GB\ data set size), and Graph500 (scale 24, edge factor 16)
    subject to active core count and CPU core frequency on SNB.  }
    \label{SNB-P_base-different-codes}
\end{figure}
In \cite{Hager:2012} a simple power dissipation model for multicore
CPUs was proposed, which assumed a baseline (static) and a dynamic power
component, with only the latter being dependent on the number of
active cores: $P=P_\mathrm{static}+nP_\mathrm{dyn}$, with
$P_\mathrm{dyn}=P_1f+P_2f^2$. Several interesting conclusions can be
drawn from this, but the agreement of the model with actual power
measurements remains unsatisfactory: The decrease of dynamic per-core
power in the saturated performance regime, per-core static power, and
the dependence of static power on the core frequency (via the
automatically adjusted supply voltage) are all neglected. Together
with the inaccuracies of the original ECM performance model,
predictions of energy consumption become cursory at
best~\cite{Wittmann:2016}.  Moreover, since the introduction of the
Uncore clock domain with the Intel Haswell processor, a single clock
speed $f$ became inadequate. An improved power model can be
constructed, however, by adjusting some assumptions:
\begin{itemize}
\item There is a baseline (``static'') power component for the whole chip
  (i.e., independent of the number of active cores) that is not
  constant but depends on the clock speed of the Uncore:\footnote{Although
    Sandy Bridge and Ivy Bridge processors do not have a
    separate Uncore frequency domain, on chips based on these
    microarchitectures the Uncore is clocked at the same speed as the cores, which
    implies a variate Uncore frequency and thereby a non-constant baseline
    power on these processors as well.}
  \bq\label{eq:base}
  P_\mathrm{base}(f_\mathrm{Uncore}) = W_0^\mathrm{base}+W_1^\mathrm{base}f_\mathrm{Uncore}
  + W_2^\mathrm{base}f_\mathrm{Uncore}^2\eos
  \eq
\item As long as there is no bandwidth bottleneck there is a power
  component per active core, comprising static and dynamic power
  contributions:
  \bq\label{eq:core}
  P_\mathrm{core}(f_\mathrm{core},n)=W_0^\mathrm{core}+\left(W_1^\mathrm{core}f_\mathrm{core}
  + W_2^\mathrm{core}f_\mathrm{core}^2\right)\varepsilon(n)^\alpha\eos
  \eq
  In the presence of a bandwidth bottleneck, performance stagnates
  but power increases (albeit more slowly than in the scalable case)
  as the number of active cores goes up. We accommodate this behavior
  by using a damping factor $\varepsilon(n)^\alpha$, where $\varepsilon(n)$
  is the parallel efficiency at $n$ cores and $\alpha$ is a fitting
  parameter. 
\end{itemize}
The complete power model for $n$ active cores is then
\bq\label{eq:full}
P_\mathrm{chip} = P_\mathrm{base}(f_\mathrm{Uncore}) + n P_\mathrm{core}(f_\mathrm{core},n)\eos
\eq
The model fixes all deficiencies of the original formulation,
but this comes at the price of a significant number of fitting
parameters. The choice of a quadratic polynomial for the $f$
dependence is to some extent arbitrary; it turns out that a cubic
term does not improve the accuracy, nor does an exponential
form such as $\beta+\gamma f^\delta$\@. 
Thus we stick to the quadratic form in the following.
Note that there is a connection between the model parameters
  and ``microscopic'' energy parameters such as the energy per cache
  line transfer, per floating-point instruction, etc., which we do not
  use here since they also result from fitting procedures; they also
  cannot predict the power dissipation of running code with sufficient
  accuracy.

\begin{figure}[tbp]
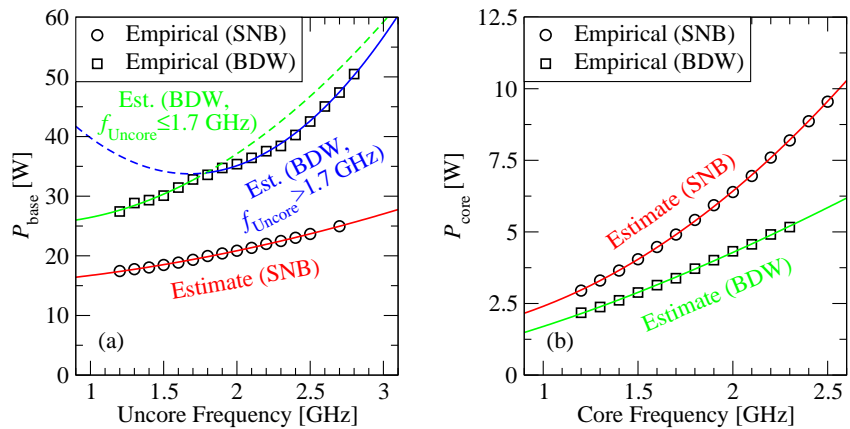

    \centering
    \hspace*{\fill}
    \includegraphics*[scale=0.75]{P_base-combined}
    \hfill
    \includegraphics*[scale=0.75]{P_core-combined}
    \hspace*{\fill}
    \caption{(a) $P_\mathrm{base}$ and (b) $P_\mathrm{core}$ parameters derived from empirical data for
      different CPU core/Uncore frequencies on SNB and BDW. The $P_\mathrm{core}$
      fit was done for the \textsc{dgemm} benchmark; \textsc{stream}
      yields different fitting parameters (see Table~\ref{tab:parameters}).}
    \label{power-model-fits}
\end{figure}
The model parameters $W_*^*$ and $\alpha$ have to be determined by
fitting procedures, running code with different characteristics.
In Figure~\ref{SNB-P_base-different-codes}(b) we show how
$P_\mathrm{base}$ is determined by extrapolating the measured power
consumption towards zero cores at different clock frequencies on SNB
(there is only one clock domain on this architecture). The
\textsc{stream} triad, a \textsc{dgemm}, and the Graph500 benchmark
were chosen because they have very different requirements towards
the hardware. In case of \textsc{stream} we ignore data points beyond
saturation (more precisely, for parallel efficiency smaller than
90\%) in order to get sensible results. The
extrapolation shows that the baseline power is independent of
the code characteristics, which is surprising since the Uncore includes
the L3 cache, whose power consumption is certainly a function of
data transfer activity. Its variation with the clock
speed can be used to determine the three parameters
$W_*^\mathrm{base}$, as shown in Figure~\ref{power-model-fits}(a)
for both architectures. In this figure, each data point (circles and
squares) is an extrapolated baseline power measurement for a different
(Un)core frequency. On the BDW CPU, the measurements exhibit a
peculiar change of trend as the frequency falls below 1.7\,\GHZ;
a different set of fit parameters is needed in this regime. We can only
speculate that the chip employs more aggressive power saving techniques
at low $f_\mathrm{Uncore}$.

\begin{table}[b]
  \centering
  \begin{tabular}{l c c}
      \hline
    Microarchitecture               & SNB               & BDW               \\
      \hline
    $\alpha$                        & 0.4               & 0.5               \\
      $W_0^\mathrm{base}$ [W]       & 14.62             &
      27.2\textsuperscript{1} \hspace{0.35em}/\hspace{0.35em}70.8\textsuperscript{2}\\
      $W_1^\mathrm{base}$ [W/\GHZ]  & 1.07              & $-6.45$\textsuperscript{1} \hspace{0.35em}/\hspace{0.35em}$-44.1$\textsuperscript{2} \\ 
      $W_2^\mathrm{base}$ [W/\GHZ$^2$]&  1.02             & 5.71\textsuperscript{1}\hspace{0.35em}/\hspace{0.35em}13.1\textsuperscript{2}\\
      $W_0^\mathrm{core}$ [W]       &  1.42 (\textsc{dgemm} / 1.33 (\textsc{stream}) &  $-0.11$ (\textsc{dgemm}) / 0.45 (\textsc{stream}) \\
    $W_1^\mathrm{core}$ [W/\GHZ]  &  $-0.52$ (\textsc{dgemm}) / 0.80 (\textsc{stream}) & $-1.46$ (\textsc{dgemm}) / 2.95 (\textsc{stream})  \\
    $W_2^\mathrm{core}$ [W/\GHZ$^2$]&  1.51 (\textsc{dgemm}) / 1.22 (\textsc{stream}) & 1.47 (\textsc{dgemm}) / $-0.24$ (\textsc{stream}) \\
      \hline
      \multicolumn{3}{l}{\textsuperscript{1}\hspace{0.1em}\footnotesize{$f_\mathrm{Uncore} \leq 1.7\,\mathrm{GHz}$}\hspace{1.40em}\textsuperscript{2}\hspace{0.1em}\footnotesize{$f_\mathrm{Uncore} > 1.7\,\mathrm{GHz}$}}
  \end{tabular}
  \caption{\label{tab:parameters}Fitted parameters for the power model
    (\ref{eq:base})--(\ref{eq:full}), using the \textsc{stream} and \textsc{dgemm}
    benchmarks. Note that these numbers are fit parameters only; their
    physical relevance should not be overstressed.}
\end{table}
As for the core power parameters $W_*^\mathrm{core}$, they do depend on the
code as can already be inferred from the data in
Figure~\ref{SNB-P_base-different-codes}(b). In Figure~\ref{power-model-fits}(b)
we show the quality of the fitting procedure for \textsc{dgemm}. The
parameter $\alpha$, which quantifies the influence of parallel efficiency
reduction on dynamic core power in the saturated regime of \textsc{stream},
can be determined as well. Table~\ref{tab:parameters} shows all fit
parameters.

\section{Energy model and validation}\label{sec:energyval}

\begin{figure}[tbp]
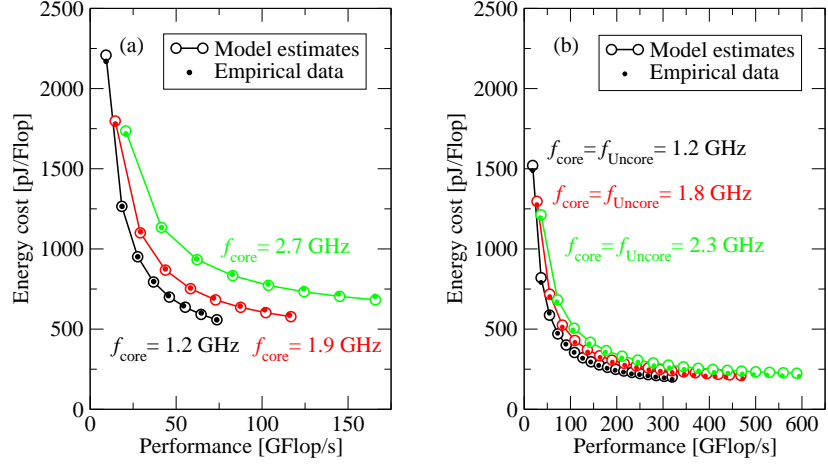

    \centering
    \hspace*{\fill}
    \includegraphics*[scale=0.70]{model-validation-Z-plot-DGEMM-phinally}
    \hfill
    \includegraphics*[scale=0.70]{model-validation-Z-plot-DGEMM-broadep2}
    \hspace*{\fill}
    \caption{Z-plots relating performance and energy cost for \textsc{dgemm}
    with from Intel's MKL on the (a) Sandy Bridge-EP processor for $N=40,000$
    and on the (b) Broadwell-EP processor for $N=60,000$ using different CPU
    core counts as well as CPU core and Uncore frequencies on the Broadwell-EP
    processor.}
    \label{model-validation-Z-plot-DGEMM}
\end{figure}
Putting together performance and power dissipation,
we can now validate predictions for energy consumption.
Guidelines for energy-efficient execution as
derived from this data will be given in Sect.~\ref{sec:consequences}
below.  We normalize the energy to the work, quantifying the energy
cost per work unit:
\bq\label{eq:emodel_full}
E = \frac{P_\mathrm{chip}(f_\mathrm{core},f_\mathrm{Uncore},n)}{\pi(f_\mathrm{core},f_\mathrm{Uncore},n)}
\eq
In our case this quantity has a unit of J/\flop.
Unfortunately, this model is  too intricate to deliver
general analytic predictions of minimum energy or EDP and the required
operating points to attain them. Some simple cases, however, can be
tackled. On a CPU with only one clock speed domain (such as SNB),
where $f_\mathrm{core}=f_\mathrm{Uncore}=f$,
and assuming that the code runtime is proportional to the inverse
clock frequency, one can differentiate (\ref{eq:emodel_full}) with
respect to $f$ and set it to zero in order to get the frequency
for minimum energy consumption. This yields
\bq\label{eq:fopt_snb}
f_\mathrm{opt}=\sqrt{\frac{W_0^\mathrm{base}+nW_0^\mathrm{core}}{W_2^\mathrm{base}+nW_2^\mathrm{core}}}\cma
\eq
which simplifies to the expression derived in~\cite{Hager:2012}
if we set $W_0^\mathrm{core}=W_2^\mathrm{base}=0$. The optimal frequency
is large when the static power components dominate, which is plausible
(``race to idle'').

We have chosen the
SNB and BDW processors for our study because they are representatives
of server CPU microarchitectures that exhibit significantly different
power consumption properties. The \textsc{dgemm} and \textsc{stream}
benchmarks are used for validation; it should be emphasized that
almost all parameters in the energy and power models (apart from
the base power) are code-dependent, so our validation makes no claim
of generality other than that it is suitable for codes with substantially
different characteristics. For \textsc{stream} we constructed
the refined ECM model as described in Sect.~\ref{sec:ecmref}, while for
\textsc{dgemm} we assumed a performance of 95\% of peak, which is
quite accurate on the two platforms. The ``Turbo'' feature was
disabled.

In order to discuss performance and power behavior the
\emph{Z-plot}~\cite{Freeh:2007,Wittmann:2016} has proven to be
useful. It is a Cartesian plot that shows (normalized or absolute)
energy consumption versus code performance,\footnote{Wallclock time can
  also be used, which essentially mirrors the plot about the $y$
  axis.} with some parameter varying along the data set. This can be,
e.g., the number of active cores, a clock frequency, a loop nest tile
size, or any other parameter that affects energy or runtime. In a
Z-plot, lines of constant energy cost are horizonal, lines of constant
performance are vertical (e.g., a \rl\ limit is a hard barrier), and
lines of constant energy-delay product (EDP) are lines through the
origin whose slope is proportional to the EDP (assuming constant
amount of work).

Figure~\ref{model-validation-Z-plot-DGEMM} shows Z-plots comparing
model predictions (open circles) and measurements (dots) for
\textsc{dgemm} on the two platforms, with varying number of active
cores along the data sets. In case of SNB
(Fig.~\ref{model-validation-Z-plot-DGEMM}(a)), each of the three data
sets is for a different core frequency (and hence implicitly
different Uncore frequency). To mimic the SNB behavior on BDW
(Fig.~\ref{model-validation-Z-plot-DGEMM}(b)), we have set the core
and Uncore frequencies to the same value.
The accuracy of the energy model is impressive, and much improved over
previous work.
\begin{figure}[tbp]
    \centering
    \includegraphics*[scale=0.65]{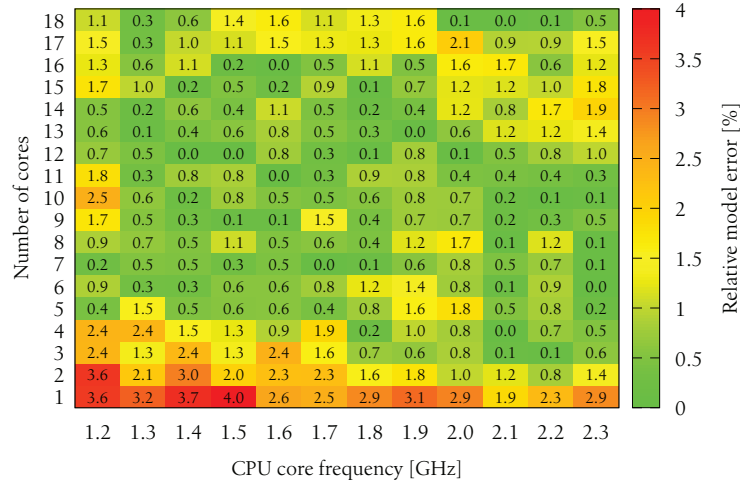}
    \caption{Relative model error for DGEMM on the Broadwell-EP processor
      for different core counts and CPU core frequencies. The Uncore
    clock speed was set to the maximum (2.8\,\GHZ).}
    \label{model-quality-BDW}
\end{figure}
As predicted by the model, lowest energy for constant
clock frequency is always achieved with all cores. The 
clock frequency for minimum energy cost at a given number of cores depends
on both parameters: the more cores,
the lower the optimal frequency due to the waning influence of the
base power. The spread in energy cost across the parameter range
is naturally larger on BDW with its large core count (18 vs.\ 8).
At full chip, both architectures show lowest EDP at the fastest
clock speed.  To get a better overview of the model quality on
BDW we show in Fig.~\ref{model-quality-BDW} a heat map of the model
error with respect to core count and core frequency at fixed Uncore
clock. The error is never larger than 4\%; if one excludes
the regions of small core counts and small core frequencies,
which are not very relevant for
practical applications anyway, then the maximum error falls below
2\% and is typically smaller than 1\%.

\begin{figure}[tb]
    \centering
    \includegraphics*[scale=0.75]{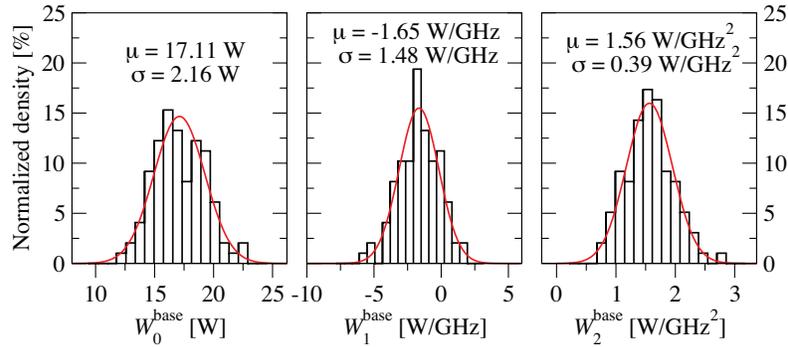}
    \caption{Histograms of $W_*^\mathrm{base}$ for \textsc{dgemm}
      among 100 Xeon Broadwell E5-2630v4 chips.
      The sum of all probabilities was normalized to one.}
    \label{meggie-W-base}
\end{figure}

\begin{figure}[tb]
    \centering
    \includegraphics*[scale=0.75]{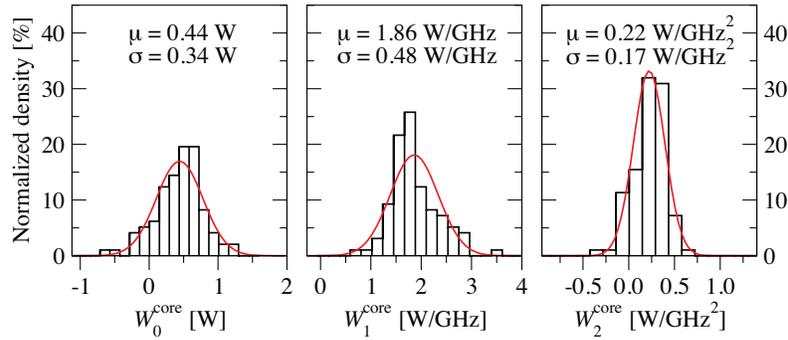}
    \caption{Histograms of $W_*^\mathrm{core}$ for \textsc{dgemm} among
      100 Xeon Broadwell E5-2630v4 chips.
      The sum of all probabilities was normalized to one.}
    \label{meggie-W-core}
\end{figure}

It is well known that manufacturing variations cause significant
fluctuation across chips of the same type in terms of power
dissipation \cite{Inadomi:2015,Hofmann:2017}.
This poses problems, e.g., when power capping is enforced
because power variations then translate into performance
variations~\cite{Inadomi:2015}, but it can also be leveraged
for saving energy by intelligent scheduling~\cite{Wilde:2015}.
For modeling purposes it is
interesting to analyze the variation of fitted power model parameters
in order to see how generic these values
are. Figures~\ref{meggie-W-base} and \ref{meggie-W-core} show
histograms of $W_*^\mathrm{base}$ and $W_*^\mathrm{core}$ for \textsc{dgemm},
including
Gaussian fits for 100 chips of a cluster\footnote{\color{red}Link to
  cluster doc hidden for double-blind review\color{black}} based on
dual-socket nodes with Intel Xeon Broadwell E5-2630v4 CPUs (10 cores).
The data clearly shows that the accuracy of the power dissipation
model can only be achieved for a particular specimen; however, the general
insights are unchanged. It cannot be ruled out that some of the
variation is caused by changes in environmental conditions across
the nodes and CPUs. For example, the typical front-to-back airflow
in many cluster node designs leads to one of the chips getting
warmer.
The (weak) bi-modal distribution of $W_0^\mathrm{base}$ may
be a consequence of this. We have also observed that chips with
a particularly high value of one parameter (e.g., $W_2^\mathrm{base}$)
are not necessarily ``hot'' specimen, because other parameters can be
average or even smaller. These observations underpin our claim
that one should not put too much physical interpretation into
the power model fit parameters but rather take them as they are
and try to reach qualitative conclusions, although the predictions
for an individual chip are accurate.

\begin{figure}[tb]
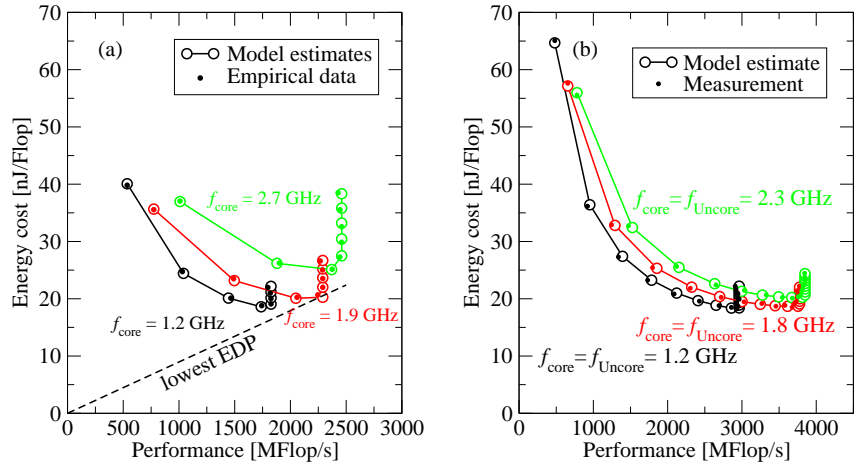

    \centering
    \hspace*{\fill}
    \includegraphics*[scale=0.70]{model-validation-Z-plot-STREAM-phinally}
    \hfill
    \includegraphics*[scale=0.70]{model-validation-Z-plot-STREAM-broadep2}
    \hspace*{\fill}
    \caption{Z-plots relating performance and energy cost for the
    \textsc{stream} triad using a 4\,GB data set for different CPU core
    counts as well as CPU core and Uncore frequencies on the (a) Sandy
    Bridge-EP and (b) Broadwell-EP processors. On BDW, core and Uncore clock
    frequencies were set to the same value for this experiment.}
    \label{model-validation-Z-plot-STREAM}
\end{figure}
In Figure~\ref{model-validation-Z-plot-STREAM} we show Z-plots comparing
the predictions and measurements for the \textsc{stream} triad with the same
frequency settings as in Fig.~\ref{model-validation-Z-plot-DGEMM}. The saturation
bandwidth, which limits the performance to the right in the plot,
was taken from the data in Fig.~\ref{SNB-P_base-different-codes}(a).
The prediction accuracy is not worse than for \textsc{dgemm}, despite
the fact that model is now much more complicated due to the saturating
performance and the drop in parallel efficiency beyond the saturation point.
A major difference between the two processors strikes the eye:
The waste in energy for core counts beyond saturation is
considerably smaller on BDW (although it is still about 20-25\%),
and the saturation point is
almost independent of the clock speed. Only the refined ECM
model can predict this accurately; in the original model,
the saturation point depends very strongly on the frequency.
At saturation, the energy consumption varies only weakly
with the clock speed, which makes finding the saturation
point the paramount energy optimization strategy. In contrast,
on SNB one has to find the saturation point \emph{and} choose
the right frequency for the global energy minimum. If the
EDP is the target metric, finding the optimal operating point
is more difficult. For both chips it coincides with the saturation
point at a frequency that is somewhere half-way between minimum and
maximum.



In summary, our power model yields meaningful estimates of high quality
with an error below 1\% for relevant operating points (i.e., away from
saturation and using more than a single core).
In contrast to the work in~\cite{Hofmann:2017}, where
the power/performance behavior was only observed empirically, we have
presented an analytic  model based on simplifying assumptions that
can accurately describe the observed behavior.


%
%
%
%
%
%
%
%


\section{Consequences for energy optimization}\label{sec:consequences}

It is satisfying that our refined ECM and power models are accurate
enough to describe the energy consumption of scalable and
bandwidth-saturating code with unprecedented quality on two quite
different multicore architectures. However, in order to go beyond an
exercise in curve fitting, we have to derive guidelines for the
efficient execution of code that are independent of the specific fit
parameters determined for a given chip specimen. As usual, we differentiate
between scalable and saturating code, exemplified by \textsc{dgemm} and
the \textsc{stream} triad, respectively.

\subsection{Scalable code}\label{sec:scalable}

\begin{figure}[tb]
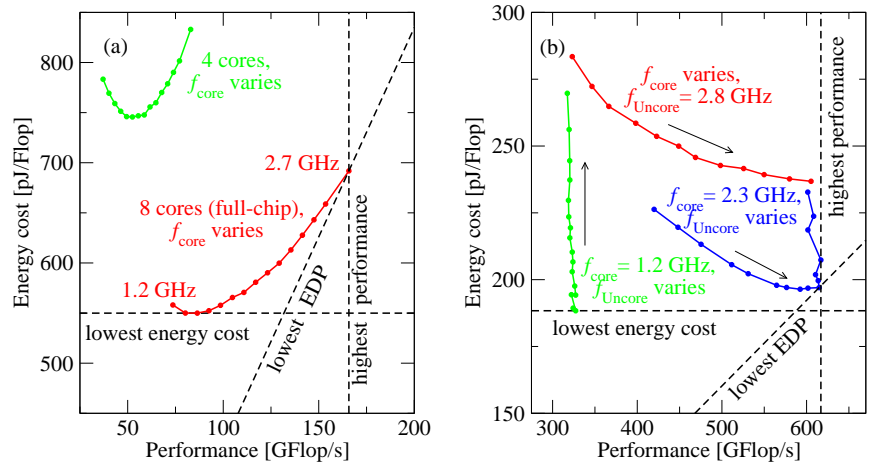

    \centering
    \hspace*{\fill}
    \includegraphics*[scale=0.70]{best-practises-Z-plot-DGEMM-SNB}
    \hfill
    \includegraphics*[scale=0.70]{best-practises-Z-plot-DGEMM-BDW}
    \hspace*{\fill}
    \caption{(a) Z-plot relating performance and energy cost for
      \textsc{dgemm} from Intel's MKL for $N=40,000$ running on four
      (half-chip) respectively eight (full-chip) cores of the Sandy
      Bridge-EP processor clocked at different CPU core frequencies,
      i.e., the core frequency varies along the curves. The energy
      minimum is exactly at the optimal frequency predicted by
      (\ref{eq:fopt_snb}). (b) Z-plot
      relating performance and energy cost for \textsc{dgemm} from
      Intel's MKL for $N=60,000$ running on all cores of the
      Broadwell-EP processor clocked at different CPU core frequencies
      and fixed, maximum Uncore clock (along the red curve) and at
      fixed maximum (blue curve) and minimum (green curve) core
      frequency with varying Uncore speed. Black arrows indicate the
      direction of rising (Un)core clock frequency.}
    \label{best-practises-Z-plot-DGEMM}
\end{figure}

Figure~\ref{best-practises-Z-plot-DGEMM}(a) shows a Z-plot with two
data sets (four and eight cores, respectively) and the core frequency
as a parameter for the SNB processor running \textsc{dgemm}.  The
highest performance, lowest energy, and lowest EDP observed are marked
with dashed lines. From the energy model and our measurements we
expect minimum energy for full-chip execution at a clock speed which
is determined by the ratio of the baseline power and the $f^2$
component of dynamic power (see (\ref{eq:fopt_snb})).  For the chip at
hand, this minimum is at $f_\mathrm{opt}\approx 1.4\,\GHZ$ with all
cores and at $f_\mathrm{opt}\approx 1.7\,\GHZ$ with only four
cores. The global performance maximum (and EDP minimum) is at the
highest core clock speed using all cores, as predicted by the model.
Hence, on this chip, where the Uncore and core frequencies are the
same by design, there is only a choice between highest performance
(and, at the same time, lowest EDP) or lowest energy consumption. The
latter is achieved at a rather low clock speed setting using all
cores. About 21\% of energy can be saved by choosing $f_\mathrm{opt}$,
albeit at the price of a 50\% performance loss.

The situation is more intricate on BDW, where the Uncore clock speed
has a strong impact on the power consumption as well as on the
performance even of \textsc{dgemm}. Figure~\ref{best-practises-Z-plot-DGEMM}(b)
shows energy-performance Z-plots for different operating modes:
Along the red curve the core clock speed is varied at maximum
Uncore clock. This is also the mode in which most production clusters are
run today since the automatic Uncore frequency scaling of the BDW processor
favors high Uncore frequencies. In this case the energy-optimal
core frequency is beyond the accessible range on this chip,
which is why the lowest-energy (and highest-performance)
``naive'' operating point is at the largest $f_\mathrm{core}$. Starting
at this point one can now reduce the Uncore clock frequency at constant,
maximum core clock (2.3\,\GHZ) until the slow Uncore
clock speed starts to impact the \textsc{dgemm} performance (blue
curve) due to the slowdown of the L3 cache.
At $f_\mathrm{Uncore}=2.1\,\GHZ$ we arrive at the global
performance maximum and EDP minimum, saving about 17\% of energy
compared to the naive operating point. At even lower  $f_\mathrm{Uncore}$
the performance starts to degrade, ultimately leading to a rise
in energy cost.
The question arises whether one could save even more energy by
accepting a performance degradation, just as on the SNB CPU. The green
curve shows the extreme case where the core clock speed is at the
minimum of 1.2\,\GHZ. Here the Uncore frequency cannot be lowered to a
point where it impacts the performance, which thus stays constant, but
the energy consumption goes down significantly. However, the additional
energy saving is only about 5\% compared to the case of maximum
performance at optimal Uncore frequency. This does not justify
the almost 50\% performance loss.  

In conclusion, the BDW CPU shows a qualitatively different
performance/energy tradeoff due to its large and power-hungry
Uncore. The Uncore clock speed is the dominating parameter here.
It is advisable to set the core clock speed to a maximum and then
lower the Uncore clock until performance starts to degrade. This
is also the point where the global EDP minimum is reached. For codes
that are insensitive to the Uncore (e.g., with purely core-bound
performance characteristics), it should be operated at the
lowest possible Uncore frequency setting.

\subsection{Saturating code}\label{sec:saturating}

A code with saturating performance characteristics due to the memory
bandwidth bottleneck is more complicated
to describe than a scalable code, because the saturation point marks
an abrupt change in the energy behavior. The saturation point (i.e.,
the number of cores required to saturate) depends on the clock speed(s) of the
chip, so it cannot be assumed that the minimum energy point is reached
at the same number of cores (as was the case for saturating behavior).

\begin{figure}[tb]
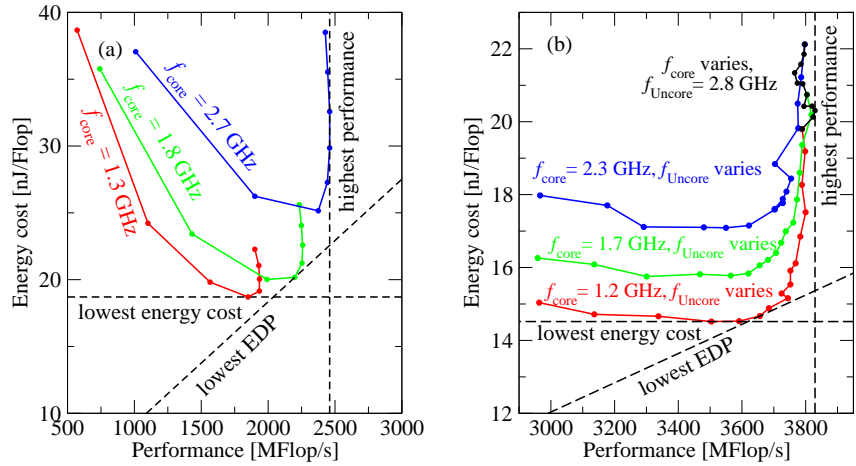

    \centering
    \hspace*{\fill}
    \includegraphics*[scale=0.70]{best-practises-Z-plot-STREAM-SNB}
    \hfill
    \includegraphics*[scale=0.70]{best-practises-Z-plot-STREAM-BDW}
    \hspace*{\fill}
    \caption{Z-plots relating performance and energy cost for the
      \textsc{stream} triad with a 4\,GB data set using various CPU
      core counts as well as core and Uncore frequencies. (a) Sandy
      Bridge-EP at three clock frequencies with varying number of
      cores along the curves. (b) Broadwell-EP, three core frequencies
      (red/green/blue), varying Uncore clock speed along the curves.
      Black: fixed Uncore frequency (at maximum), varying core clock
      speed along curve.  The number of cores at each data point on
      BDW was determined by minimizing the EDP vs.\ cores at fixed
      clock speeds. }

    \label{best-practises-Z-plot-STREAM}
\end{figure}
Figure~\ref{best-practises-Z-plot-STREAM}(a) shows Z-plots for SNB
with the \textsc{stream} triad code and three different clock speeds
(1.3, 1.8, and 2.7\,\GHZ). The number of cores varies from one to
eight along the curves. On this CPU the lowest-energy and
highest-performance operating points are quite distinct; the
saturation point with respect to core count can be clearly identified
by the lowest EDP (per core frequency), and coincides with the
highest-performance point with good accuracy. Hence, there is a simple
tradeoff between performance and energy, with a performance loss of
25\% for 28\% of energy savings (only considering the saturation
points). As mentioned before, using the whole chip is wasteful,
especially at a fast clock speed.

Already during the analysis of scalable code the Uncore clock frequency
was identified as a decisive factor on BDW.
In
Figure~\ref{best-practises-Z-plot-STREAM}(b) we thus show in red,
green, and blue the Z-plots for three different core clock speeds
(1.2, 1.7, and 2.3\,\GHZ) with the Uncore clock as the parameter along
the curves. At each point, the number of cores was determined by the minimum
EDP vs.\ active cores with fixed frequencies. As indicated in
Figure~\ref{SNB-P_base-different-codes}(a), there is a minimum Uncore
frequency required to saturate the memory interface;
Figure~\ref{best-practises-Z-plot-STREAM}(b) shows that it is largely
independent of the core frequency. In other words, there is an
$f_\mathrm{core}$-independent $f_\mathrm{Uncore}\approx 2\,\GHZ$ that
leads to (almost) maximum performance. $f_\mathrm{core}$ can then be
set very low (1.2\,\GHZ) for optimal energy and EDP without
a significant performance loss. Again it can be observed that
a sensible setting of the Uncore frequency is the major
contributor to saving energy on BDW. The black data set
shows a typical operating mode in practice, where the Uncore
clock is set very high (or left to be determined by Uncore
frequency scaling) and the core clock is varied. Even with
sensible concurrency throttling, the energy cost only
varies by about 10\%, whereas optimal parameters allow
for additional savings between 27\% and 33\%.



\section{Summary and outlook}\label{sec:summary}

By refining known ECM performance and power models we have constructed
an analytic energy model for Intel multicore CPUs that can predict the energy
consumption and runtime of scalable and saturating loop code with high
accuracy. The power model
parameters show significant manufacturing variation among CPU specimen.
The Uncore frequency on the latest Intel
x86 designs was identified as a major factor in energy optimization,
even more important than the core frequency, for scalable and saturating
code alike. Overall, energy savings of 20-50\%
are possible depending
on the CPU and code type by choosing the optimal operating point
in terms of clock speed(s) and number of active cores. If the energy-delay
product (EDP) is the target metric, the Z-plot delivers a simple yet
sufficiently accurate method to determine the point of lowest EDP.

Our work can be extended in several directions. The refined ECM model
should be tested against a variety of codes to check the generality of
the recursive latency penalty. We have ignored the ``Turbo Mode''
feature of the Intel CPUs, but our models should be able to encompass
Turbo Mode if the dynamic clock frequency variation (depending mainly
on the number of active cores) is properly taken into account. A
related problem, the reduction of the base clock speed when using AVX
SIMD instructions on the latest Intel CPUs, could be handled
in the same way. An analysis of the new Intel Skylake-SP and AMD Epyc
server CPUs for their performance and power properties is
currently ongoing. It would furthermore be desirable to identify
more cases where the energy model (\ref{eq:emodel_full}) can yield
simple analytic results. Finally, it should be useful to ease the
construction of our improved analytic performance and energy models by
extending tools such as Kerncraft~\cite{Hammer:2017}.



\end{document}